\documentclass[a4paper]{iopart}
\usepackage{graphicx}

\newcommand\rs[1]{{\scriptscriptstyle\rm #1}}
\newcommand\text[1]{{\rm #1}}

\begin{document}

\title{Quantum manipulation in a Josephson LED}

\author{Fabian Hassler$^{1,2}$, Yuli V Nazarov$^2$ and Leo~P~Kouwenhoven$^2$}

\address{$^1$ Instituut-Lorentz, Universiteit Leiden, P.O. Box 9506, 2300 RA
Leiden, The Netherlands}

\address{$^2$ Kavli Institute of Nanoscience, Delft University of Technology, P.O.
Box 5046, 2600 GA Delft, The Netherlands}

\ead{Y.V.Nazarov@tudelft.nl}

\begin{abstract}
We access the suitability of the recently proposed Josephson LED for
quantum manipulation purposes. We show that the device can both be used for
on-demand production of entangled photon pairs and operated as a two-qubit
gate. Besides, one can entangle particle spin with photon polarization
and/or measure the spin by measuring the polarization.
\end{abstract}

\section{Introduction}

It is tempting to use the advantages of semiconductors and superconductors,
combined within a single nanodevice, for quantum manipulation purposes.
Making such combined nanostructures turned out to be a difficult
technological problem and a lot of experimental effort has been
concentrated on this direction~\cite{vanwees:97}.  Progress has been
achieved with semiconductor nanowires: Superconducting field-effect
transistor~\cite{doh:05} and Josephson effect~\cite{vandam:06,xiang:06} in a
semiconducting quantum dot have been experimentally confirmed.  Recently,
a next step has been made. It was proposed to combine semiconducting
quantum dots and superconducting leads to make a Josephson LED where the
light-emission ability of a semiconductor is enhanced by the intrinsic
coherence of the superconducting state \cite{recher:09}.

Semiconducting quantum dots exhibit narrow emission lines and
quasi-atomic discrete states, this enables quantum applications involving
visible photons.  The optical emission shows close to perfect photon
antibunching \cite{michler:00,zwiller:01,santori:02}, so the dots can
be used as single-photon emitters. Rabi oscillations \cite{zrenner:02}
and coherent manipulation of excitons (electron-hole bound states)
have been demonstrated \cite{li:03}. Furthermore, the possibility
of controlled charging with extra carriers \cite{warburton:00}
allows the use of single electron \cite{atature:06,atature:07} or hole
\cite{gerardot:08,brunner:09} spins that exhibit ultra-long spin-coherence
times \cite{khaetskii:00}. Importantly, biexciton cascades, which are sources
of photon pairs emitted sequentially \cite{moreau:01}, were proposed to
generate polarization entangled photons \cite{benson:00,gywat:02}. However,
the exchange splitting of a single exciton due to asymmetric dots renders
the two possible circular polarizations nondegenerate and hinders the
observation of entanglement \cite{santori:02a}. This problem can be
overcome by improvements in the sample design \cite{stevenson:06} or by
spectral filtering \cite{akopian:06}. A more serious disadvantage is the
use of incoherent transitions to prepare a biexcitation. Owing to this,
it is hard if possible at all to generate entangled pairs on demand,
a functionality that is required in most quantum algorithms \cite{nielsen}.

In this article, we address the rich potential of the newly proposed
Josephson LED for quantum manipulation purposes. We show how to operate the
device for on-demand production of entangled photon pairs. We demonstrate
that Josephson LEDs may be used as a two-qubit quantum gates.  Moreover,
we show how to entangle the spin of a particle in one of the quantum dots
with the polarization of an emitted photon.  We also outline an alternative
scheme to measure the spin of the particle via the conversion of the spin
into the polarization of a photon.

\section{Setup}

\begin{figure}
  \centering
  \includegraphics{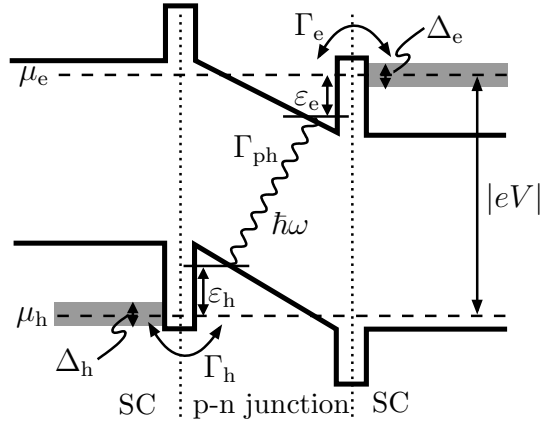}
  \caption{%
  Level diagram of the double quantum dot in a p-n junction where
  each side is coupled to a superconducting lead with pairing amplitude
  $\Delta_\text{h,e}$.  The levels are detunes by $\varepsilon_\text{h,e}$
  with respect to the chemical potential $\mu_\text{h,e}$ on either side.
  The coupling to the superconducting leads introduces coherent transfer
  of Cooper pairs changing the occupation number of the dot by 2.  The p-n
  junction is biased with a voltage $V$ which sets the energy scale of
  electron-hole recombination via the emission of photons at frequency
  $\omega= |eV|/\hbar$.
  }\label{fig:setup}
\end{figure}
The setup of the Josephson light emitting diode (JoLED) was outlined
in detail in~\cite{recher:09}.  It consists of a p-n junction in a
semiconducting wire where either side features a quantum dot. Each
quantum dot can incorporate up to two holes (h) or electrons (e) in a
single level. The potential barriers are arranged to assure that the only
process of charge transfer through the junction is the recombination of
an electron and a hole in the dots.  The p-n junction is biased with a
voltage $V$. This sets the energy scale $|eV|$ for photons emitted via
recombination of a electron and a hole at a rate $\Gamma_\text{ph}$.
Either dot is coupled to a superconducting lead with a pairing amplitude
$\Delta_\text{h,e} = |\Delta_\text{h,e}| \exp(i\phi_\text{h,e})$. Thereby,
each superconducting lead introduces mixing between the empty and the
doubly occupied state of the dots via the proximity effect; we denote by
$\tilde \Delta_\text{h,e} = (\Gamma_\text{h,e} /2) \exp(i\phi_\text{h,e})$
the induced paring amplitude on the dots with $\Gamma_\text{h,e}$ the level
broadening proportional to the square of the amplitude to tunnel an electron
from the dot to the superconducting lead.  The Hamiltonian $H = H_\text{d} +
H_\text{m}$ of the system consists of two terms.  The first term
\begin{equation}\label{eq:charging}
  H_\text{d} =  \varepsilon_\text{h} n_\text{h} + U_\text{h} n_\text{h}
  (n_\text{h} -1) + \varepsilon_\text{e} n_\text{e} + U_\text{e} n_\text{e}
  (n_\text{e} -1) + U_\text{he} n_\text{h} n_\text{e}
\end{equation}
is diagonal in the charge basis with $n_\text{h}$ ($n_\text{e}$)
being the number of holes (electrons); here, $\varepsilon_\text{h}$
($\varepsilon_\text{e}$) denotes the level of the dot with respect to the
chemical potential $\mu_\text{h}$ ($\mu_\text{e}$) in the p (n) region,
$U_\text{h,e}$ denotes the on-site charging energy and $U_\text{he}$ is
the Coulomb interaction between the carriers in the dots.  The effect of
$U_\text{he}$ has not been considered in \cite{recher:09} and is an important
detail of our setup.  The second term (due to the superconducting leads)
\begin{equation}\label{eq:mixing}
  H_\text{m} = \tilde\Delta_\text{h} |2_\rs{h} \rangle \langle 0_\rs{h} | +
   \tilde\Delta_\text{e} |2_\rs{e} \rangle \langle 0_\rs{e} | + \text{H.c.},
\end{equation}
introduces mixing between states with well-defined charge; here and in the
following, $| n_\rs{h} \rangle$ ($| n_\rs{e} \rangle$) denotes the state
with n holes (electrons) on the left (right) dot.  In \cite{recher:09},
a general case $\tilde{\Delta} \simeq U$ has been considered so that
the mixing between the charge states has been always essential. Here,
we are interested in a limit where $\tilde \Delta \ll U \ll \Delta$,
i.e., expression (\ref{eq:charging}) is typically the dominant term in
the Hamiltonian and (\ref{eq:mixing}) constitutes a perturbation.

In this limit, quantum manipulation functionality is enabled. The
Hamiltonian $H_\text{d}$ without mixing naturally constitutes a two qubit
system with the four states given by $|0_\rs{h} 0_\rs{e}\rangle$, $|2_\rs{h}
0_\rs{e}\rangle$, $|0_\rs{h} 2_\rs{e}\rangle$, $|2_\rs{h} 2_\rs{e}\rangle$,
and the two qubits correspond to the two different dots.  We note that the
qubits interact with each other, since $U_{he}$ is nonzero.  For instance,
the energy difference between $|0_\rs{e}\rangle$ and $|2\rs{e}\rangle$
depends on number of holes in the neighboring hole dot.  This offers the
possibility to operate two qubit gates \cite{lloyd:96}.

\section{Dynamics without manipulation}\label{sec:init}

\begin{figure}
  \centering
  \includegraphics{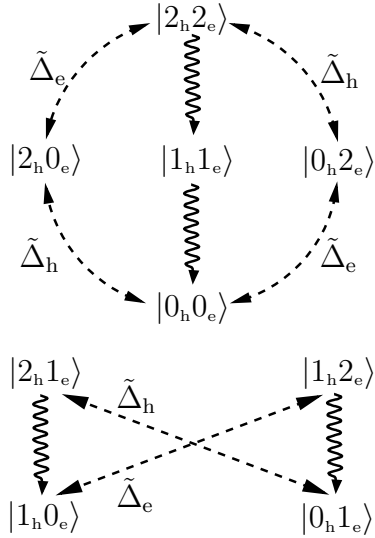}
  \caption{%
  Scheme of the 9 states (without spin degree of freedom) of the double
  quantum dot and possible transitions due to electron-hole recombination
  (wavy lines).  As the recombination annihilates one electron with one
  hole these processes cannot change the parity of the total number of
  electrons and holes on the dots.  Therefore, the diagram is separated
  into two parts: At the top, there are the even parity states and on the
  bottom the odd ones.  The dashed lines indicate possible two particle
  transfers which can happen due to the nearby superconductors.  Note that
  these processes only contribute with an appreciable probability when
  the involved states are nearly degenerate in energy.
  }\label{fig:photon}
\end{figure}
At first, we describe the dynamics of JoLED without manipulation.  As noted
above, the mixing of different charge state is small and thus we neglect
it at the moment.  We will comment on the effect of mixing at the end of
the section.  The coupling of the states of the dot to the radiation field
leads to emission of photons with frequency $\omega \approx |eV| /\hbar$.
The system we have in mind is a III-V semiconductor, e.g., GaAs or InAs,
where the electrons in the conduction band carry a spin $\pm 1/2$ while
the holes in the valence band carry a total angular momentum $3/2$.
Following the basic assumptions about spin-polarization conversion in
quantum wells, we take for granted that the angular momentum and spin of
a dot state are in the same direction \cite{weisbuch}.  This gives the
following selection rule: The recombination is only possible for electron
and hole of the same spin ($\uparrow$ or $\downarrow$) and produces a
photon of circular polarization corresponding to the direction of this
spin ($+$ or $-$).  The recombination of the states $| {\uparrow}_\rs{h}
{\downarrow}_\rs{e} \rangle$ and $| {\downarrow}_\rs{h} {\uparrow}_\rs{e}
\rangle$ is forbidden by the selection rule and happens with a much smaller
rate.  Assuming an appropriate spin configuration, the decay channels are
depicted by wavy lines in figure~\ref{fig:photon}.  We note that the parity
of the total number of electrons and holes is conserved by the process
of photon emission: The parity is even in the top and odd in the bottom
cycle in the figure.  If initially the JoLED is in the state $|2_\rs{h}
2_\rs{e}\rangle$ or $|1_\rs{h} 1_\rs{e}\rangle$, it will be in the state
$|0_\rs{h} 0_\rs{e}\rangle$ after a time $\simeq \Gamma_\text{ph}^{-1}$.
Similarly, it will be in $|0_\rs{h} 1_\rs{e}\rangle$ or $|1_\rs{h}
0_\rs{e}\rangle$ if the initial parity is odd.

There are secondary (slow) processes not depicted in figure~\ref{fig:photon}
that change the parity.  These processes emit photons together with the
creation of a quasiparticle in one of the superconducting leads and therefore
connect the even- and odd-parity cycles in figure~\ref{fig:photon}. As an
example, consider the case where the initial state is given by $|1_\rs{h}
0_\rs{e}\rangle$.  In a virtual process, an electron can tunnel in
from the superconductor on the electron side such that the dots are
now in the (intermediate) state $|1_\rs{h} 1_\rs{e}\rangle$ leaving
behind a quasiparticle with energy larger than $|\Delta_\text{e}|$
in the lead, followed by the emission of a photon such that the dots
end up in the state $|0_\rs{h} 0_\rs{e}\rangle$.  The (typical) rate
for this secondary emission is given by $\tilde \Gamma_\text{ph;e}
\simeq \Gamma_\text{ph} \Gamma_\text{e}/|\Delta_\text{e}|$ which
is smaller by $|\tilde \Delta_\text{e} / \Delta_\text{e}| \ll 1$
than the primary emission; a similar process going from $|0_\rs{h}
1_\rs{e}\rangle$ to $|0_\rs{h}0_\rs{e}\rangle$ via the creation
of a quasihole in the superconducting lead on the p-side has a
typical rate $\tilde \Gamma_\text{ph; h} \simeq \Gamma_\text{ph}
\Gamma_\text{h}/|\Delta_\text{h}|$.  Taking both the primary and secondary
photon emission processes into account, we come to the following conclusion:
The JoLED will end up in the ground state $|0_\rs{h}0_\rs{e}\rangle$ after
time $\simeq \Gamma_\text{ph;e,h}^{-1}$. This proves that the two-quit gate
is automatically prepared in the initial state $|0_\rs{h}0_\rs{e}\rangle$.
The effect of a small nonvanishing mixing $\tilde\Delta$ is now
easily discussed resorting to perturbation theory. In fact the state
$|0_\rs{h}0_\rs{e}\rangle$ is not an eigenstate of the system and the
true ground state has also components $|2_\rs{h}0_\rs{e}\rangle$,
$|0_\rs{h}2_\rs{e}\rangle$, and $|2_\rs{h}2_\rs{e}\rangle$
admixed. Those states however are generically detuned from the state
$|0_\rs{h}0_\rs{e}\rangle$ by $U$. Therefore, the amplitude to be in
state $|2_\rs{h}2_\rs{e}\rangle$ is given by $\tilde \Delta_\text{e}
\tilde\Delta_\text{h}/U^2$ in second order perturbation theory in
$H_\text{m}$ and the probability to be in state $|2_\rs{h}2_\rs{h}\rangle$
which can decay via the recombination of excitons reads $|\tilde
\Delta_\text{e}|^2 |\tilde\Delta_\text{h}|^2/U^4 \ll 1$. For that reason,
including mixing thus does not change our conclusion. The system remains in
the ground state $|0_\rs{h}0_\rs{e}\rangle$ with overwhelming probability.

\section{On-demand production of photon pairs}

So far, we were only considering the states of the Hamiltonian
(\ref{eq:charging}) together with the coupling to the radiation field.
In a next step, we introduce mixing given by (\ref{eq:mixing}). Mixing
provides a coherent coupling between the eigenstates of $H_\text{q}$
depicted by dashed lines in figure~\ref{fig:photon}.  At first, we are
interested in the case where figure~\ref{fig:photon} contains a closed
cycle such that a constant stream of photons is produced.  This can be
achieved by tuning the on-site energies $\varepsilon_\text{h,e}$ via
voltages of close-by gates.  Having two tuning parameters, we can activate
two mixing processes by tuning the relevant eigenstates into degeneracy.
We know that $|0_\rs{h} 0_\rs{e} \rangle$ is the equilibrium state without
mixing.  If we therefore tune this level into degeneracy with $|0_\rs{h}
2_\rs{e} \rangle$ and $|2_\rs{h} 2_\rs{e} \rangle$, the cycle $|0_\rs{h}
0_\rs{e} \rangle \to |0_\rs{h} 2_\rs{e} \rangle \to |2_\rs{h} 2_\rs{e}
\rangle \to |1_\rs{h} 1_\rs{e} \rangle \to |0_\rs{h} 0_\rs{e} \rangle$
becomes active in which two photons are produced with frequencies $\hbar
\omega \approx e V$.\footnote{Alternatively, we may tune $|0_\rs{h} 0_\rs{e}
\rangle$, $|2_\rs{h} 0_\rs{e} \rangle$, and $|2_\rs{h} 2_\rs{e} \rangle$
into degeneracy.} This cycle is interrupted from time to time by a secondary
photon emission which brings the system to the odd (bottom) cycle.  There it
remains for some time in one of the states $|1_\rs{h} 0_\rs{e} \rangle$
or $|0_\rs{h} 1_\rs{e} \rangle$ until the secondary photon emission brings
it back to $|0_\rs{h} 0_\rs{e} \rangle$.  The degeneracy of the states
can be obtained by setting the on-site energy levels to
\begin{equation}\label{eq:deg}
  \varepsilon^*_\text{h} = -U_\text{h} -
    2 U_\text{he} 
    \qquad
  \varepsilon^*_\text{e} = -U_\text{e}. 
\end{equation}
\begin{figure}
  \centering
  \includegraphics{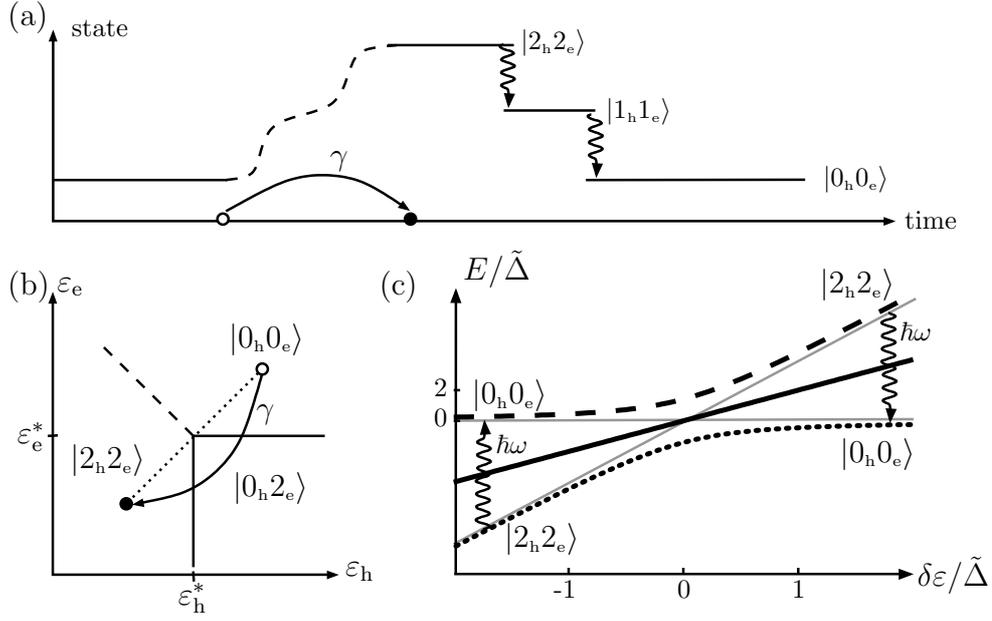}
  \caption{%
  (a) Time-line depicting the pumping of the quantum dots (between the white
  and the black dot) together with the subsequent emission of two (entangled)
  photons.  (b) Diagram showing which of the states $|0_\rs{h} 0_\rs{e}
  \rangle$, $|0_\rs{h} 2_\rs{e} \rangle$, $|2_\rs{h} 2_\rs{e} \rangle$
  has the lowest energy given a set of parameters $\varepsilon_\text{h},
  \varepsilon_\text{e}$.  At the boundaries (solid and dashed lines),
  the bordering states are degenerate neglecting the mixing introduced by
  $H_\text{m}$.  The mixing induces an energy gap along the solid lines.
  Changing the state adiabatically from the white dot to the black dot
  along the path $\gamma$, we transfer the dot from the state $|0_\rs{h}
  0_\rs{e} \rangle$ to $|2_\rs{h} 2_\rs{e} \rangle$.  (c) Level scheme
  depicting the anticrossing of the three states along the dotted
  line in (a) where $\delta \varepsilon = \delta \varepsilon_\text{h} =
  \delta\varepsilon_\text{e}$ and $\tilde \Delta= |\tilde \Delta_\text{h}|
  = |\tilde \Delta_\text{e}|$.  The cycle produces photon pairs on demand
  and is described in detail in the main text.  We start with the dots
  in the ground state $|0_\rs{h} 0_\rs{e}\rangle$ at $\delta \varepsilon
  > \tilde \Delta$.  Adiabatically changing $\delta \varepsilon$ to
  $\delta \varepsilon < - \tilde \Delta$, we drive the levels through the
  anticrossing and end up with the state $|2_\rs{h} 2_\rs{e} \rangle$ which
  relaxes to $|0_\rs{h} 0_\rs{e}\rangle$ via the emission of an entangled
  pair of photons.  Changing $\delta\varepsilon$ adiabatically back to the
  original situation, an additional pair of photons is produced leading
  to a total of two photon pairs per cycle in the $\varepsilon_\text{h,e}$
  parameter space.
  }\label{fig:phase}
\end{figure}

Close to the degeneracy point $\varepsilon_{h,e}^*$, the states $|0_\rs{h}
0_\rs{e} \rangle$, $|0_\rs{h} 2_\rs{e} \rangle$, and $|2_\rs{h} 2_\rs{e}
\rangle$ are almost degenerate and the remaining 6 states are separated by the
interaction energy $U$.  The induced superconducting gaps lead to mixing of
the dot states which can be used to excite the dot followed by photon
emission.  Denoting the detuning from the degeneracy point by $\delta
\varepsilon_\text{h,e} = \varepsilon_\text{h,e} - \varepsilon_\text{h,e}^*$, the
Hamiltonian
\begin{equation}\label{eq:red_ham}
  H'= \left(
  \begin{array}{ccc}
    0             & \tilde\Delta_\text{e}^*              & 0 \\
    \tilde\Delta_\text{e} & 2\,\delta \varepsilon_\text{e}  
    & \tilde\Delta_\text{h}^*\\
    0             & \tilde\Delta_\text{h}      
    & 2\,\delta \varepsilon_\text{h} + 2\,\delta \varepsilon_\text{e} 
  \end{array}
  \right)
\end{equation}
is almost degenerate in the subspace $\{|0_\rs{h} 0_\rs{e} \rangle, |0_\rs{h}
2_\rs{e} \rangle, |2_\rs{h} 2_\rs{e} \rangle \}$.  Figure~\ref{fig:phase}(b)
shows the parameter space $\varepsilon_\text{h}$, $\varepsilon_\text{e}$
together with the state which have to lowest energy for these
parameters.  At the boundaries, two of the states become degenerate.
Along the solid line a gap opens due to mixing of the states caused
by the superconductor so the level crossing becomes an anticrossing
gapped by $|\tilde\Delta|$.  Along the dashed line, no gap opens as the
states involved are not coupled by the Hamiltonian (\ref{eq:red_ham}).
Starting from the ground state $|0_\rs{h}0_\rs{e} \rangle$ [denoted
by the white dot in figure~\ref{fig:phase}(b)] in the region $\delta
e_\text{h} > -\delta_\text{e}, \delta_\text{e} > 0$ and moving the state
adiabatically along $\gamma$ via the state $|0_\rs{h} 2_\rs{e} \rangle$
to the black dot in the region where $|2_\rs{h} 2_\rs{e}\rangle$ is the
lowest state of $H'$, we end up with the state $|2_\rs{h} 2_\rs{e}\rangle$
which will subsequently decay via the emission of two photons, cf.\
figure~\ref{fig:photon}.  Figure~\ref{fig:phase}(c) shows the level scheme
for the case when the states are tuned through the triple point along the
dotted line in figure~\ref{fig:phase}(b).  The spectrum for paths which do
go directly through the triple point are similar but feature two instead of
one anticrossing.  After the emission of the photons, we are back in the
state $|0_\rs{h} 0_\rs{e} \rangle$ which can be repumped into $|2_\rs{h}
2_\rs{e} \rangle$ by retracing the path $\gamma$ thus completing the cycle.
Note that the pumping should be slow in order to be adiabatic but fast
such that the intermediate state $|0_\rs{h} 2_\rs{e} \rangle$ does not
decay due to secondary photon emission; this approximately translates
into $ \tilde \Delta^{-1} \ll t_\text{pump} \ll \tilde \Gamma_\text{ph}$
with $t_\text{pump}$ the pumping time.  Photons pairs can be produced at
will by employing the adiabatic pumping.  However, the photon emission
process is stochastic in its nature and the exact time when the photons
are produced cannot be controlled.  Pumping the system, we obtain a pair
of photons somewhere within the time $\Gamma_\text{ph}^{-1}$.

\begin{figure}  
  \centering
  \includegraphics{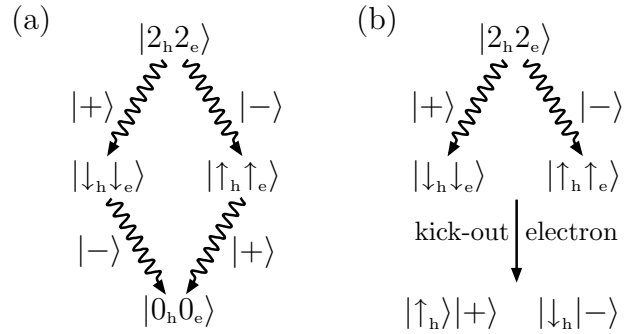}
  \caption{%
  (a) Biexciton cascade which leads to the generation of a pair of entangled
  photons. (b) For the production of spin-polarization entanglement, the
  process in (a) is interrupted after the first photon emission and the
  electron remaining in the dot is kicked out into the lead by a pulse in
  gate voltage $V_\text{e}$.
  }\label{fig:entan}
\end{figure}

The photons created in the cycle are entangled in their polarization
degree on freedom.  Starting with the state $\Psi_0=|2_\rs{h} 2_\rs{e}
\rangle$ of the dot immediately after pumping, the first photon which is
emitted can either be $+$ or $-$ polarized.  In fact, the state $\Psi_1$
after the first emission is a linear superposition of the photon being
in state $|+_\rs{1}\rangle$ or $|-_\rs{1}\rangle$ with the same amplitude
for both.  After the first photon emission, the state of the system (dot
and photon) reads
\begin{equation}\label{eq:photon1}
  \Psi_1 
  \propto 
  |{\downarrow}_\rs{h} {\downarrow}_\rs{e} \rangle |+_\rs{1}\rangle +
  |{\uparrow}_\rs{h} {\uparrow}_\rs{e} \rangle |-_\rs{1}\rangle.
\end{equation}
Note that the polarization of the photon is connected to the state of the
remaining hole and electron.  Therefore, the polarization of the photon
produced in the second recombination is linked to the polarization of the
first photon and we end up with the state
\begin{equation}\label{eq:photon2}
  \Psi_2 \propto |0_\rs{h} 0_\rs{e} \rangle 
  ( |+_\rs{1} -_\rs{2} \rangle + |-_\rs{1} +_\rs{2} \rangle ),
\end{equation}
with the polarization degrees of freedom are (completely) entangled. The
physics behind the polarization entanglement is the same as observed in
biexciton cascade in semiconducting quantum dots without superconducting
leads \cite{stevenson:06,akopian:06}. However, the pumping scheme differs as
the biexciton is electrostatically pumped without the need of an radiation
field whereas traditionally the biexciton state $|2_\rs{h} 2_\rs{e}
\rangle$ is pumped with lasers via the single exciton state $|1_\rs{h}
1_\rs{e} \rangle$ which is unstable itself and can decay such that the
pumping has to be faster than the exciton decay time.  The electrostatic
pumping may ease the detection of the entangled photons as no background
laser field is present.

\section{Qubit Manipulation}

\begin{figure}
  \centering
  \includegraphics{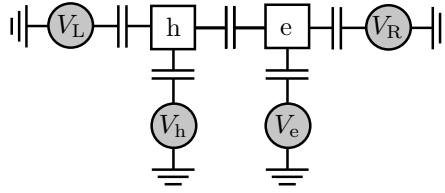}
  \caption{%
  Capacitance network showing the two dots (h and e) connected to the
  superconducting leads ($V_\text{L}$ and $V_\text{R}$) as well as to gate
  electrodes ($V_\text{h}$ and $V_\text{e}$).
  }\label{fig:capacitances}
\end{figure}

As mentioned above, the states $|0_\rs{h} 0_\rs{e}\rangle$, $|2_\rs{h}
0_\rs{e}\rangle$, $|0_\rs{h} 2_\rs{e}\rangle$, and $|2_\rs{h}
2_\rs{e}\rangle$ represent a two qubit system. In section~\ref{sec:init},
we have shown that the system left alone relaxes to the state $|0_\rs{h}
0_\rs{e}\rangle$. This provides an automatic initialization of the
qubit. In this section, we outline a possible scheme for the qubit
manipulation by irradiation pulses. It is important to note that the
manipulation is all-electric, achievable by modulating the gate voltages.
Figure~\ref{fig:capacitances} depicts the system of the two dots together
with four gates. Two of them, $V_\text{L}$ and $V_\text{R}$, address the
superconducting leads and the remaining two, $V_\text{h}$ and $V_\text{e}$,
are the back gates for the dots in the superconducting wire already used
above to tune the dots.

Since all the energy differences between the qubit states are nondegenerate,
a specific transition can be addressed by tuning the irradiation frequency
$\omega$ to the energy difference between the states involved.  To understand
the details of the manipulation, it is important to note that the voltages
not only shift the positions of the levels $\varepsilon_\text{h,e}$ with
respect to corresponding electrodes, they also produce time-shifts in the
superconducting phases of the electrodes, so that $\tilde \Delta_\text{e}
\to \tilde\Delta_\text{e} e^{i\phi_\text{R}(t)}$ with $\dot\phi_\text{R} =
2e V_\text{R}(t)/\hbar$, and similar for $\tilde \Delta_\text{h}$. Neglecting
this would lead to the confusing and incorrect conclusion that transitions
can be induced even without capacitances between the dots and the gate
electrodes. In fact, the division of the a.c.~voltage in a capacitative
network is crucial for the transitions to occur.
 
To make this explicit, it is constructive to perform a unitary
(gauge) transformation that cancels the time dependence of
$\tilde\Delta_\text{h,e}$. After this, the only effect of the a.c.~voltage
is the modulation of the levels given by
\begin{eqnarray}
\delta \varepsilon_\text{h}/e 
=  a_\text{hh} (V_\text{h} -V_\text{L}) + a_\text{he} (V_\text{e} -V_\text{L}) 
+ a_\text{h}(V_\text{R}-V_\text{L}), \\
\delta \varepsilon_\text{e}/e 
=  a_\text{ee} (V_\text{e} -V_\text{R}) 
+ a_\text{eh} (V_\text{h} -V_\text{R}) + a_\text{e}(V_\text{L}-V_\text{R});
\end{eqnarray}
where the coefficients $a$, $|a|<1$, are obtained from the voltage division
in the capacitance network.  In zeroth order in $|\tilde\Delta|/\hbar
\omega$, the irradiation pulses do not induce transitions between the qubit
states but rather change their mutual energy differences.  Transitions
appear in first order with a corresponding non-diagonal matrix element of
the order of $|\tilde{\Delta} \, \delta\varepsilon|/\hbar\omega$.

As a concrete example, let us tune $\hbar\omega$ to the energy
difference between the states $|2_\rs{h} 0_\rs{e}\rangle$ and $|2_\rs{h}
2_\rs{e}\rangle$. The Hamiltonian $H''$ in the relevant subspace spanned
by these two states reads
\begin{equation}\label{eq:ham}
  H''= \left(
  \begin{array}{cc}
    0 & \tilde \Delta_\text{e}^* \\
    \tilde \Delta_\text{e} & \hbar\omega+ 2\, \delta\varepsilon_\text{e}(t)
   \end{array}
   \right)
\end{equation}
where the energies are measured with respect to the reference
state $|2_\rs{h} 0_\rs{e}\rangle$.  A constant resonant irradiation
modulating $\delta\varepsilon_{e}(t)$ harmonically with amplitude
$\bar{\varepsilon}$ results in Rabi oscillations between these
states at a frequency $\omega_\text{R}= |\tilde \Delta_\text{e}|
J_1(2\bar{\varepsilon}/\hbar\omega)/\hbar$, where $J_1(x)$ denotes a Bessel
function of the first kind. If one applies the irradiation for a time $t_\pi=
\pi/\omega_\text{R}$, corresponding to a $\pi$ pulse, the effect is a c-NOT
gate: Depending on the state $|0_\rs{h}\rangle$ or $|2_\rs{h}\rangle$ of the
hole qubit, the state of the electron qubit is inverted or not. Note that
the c-NOT gate is a fundamental two qubit gate which together with arbitrary
single qubit operations can simulate any quantum circuit \cite{nielsen}.

The readout of the two-qubit gate occurs via the radiative decay of the
state $|2_\rs{h} 2_\rs{e}\rangle$ which is the only one with a sizable
radiative decay rate. The fact that one can read only the probability of
this state is known to present no principal obstacle for measuring more
complicated variables, since one can perform an arbitrary unitary operation
in the Hilbert space before the read-out.  In fact, full tomography of
a two qubit density matrix has been demonstrated recently by resorting
only to the measurement of a single fixed operator \cite{dicarlo:09}. We
remark that the main source of decoherence in the qubits is due to voltage
fluctuations in the environment.

\section{Photon-spin entanglement}

The state $\Psi_1$, after the emission of the first photon, exhibits
entanglement between the photon and the dot degrees of freedom.  This state
is however not stable and will decay further as explained above.  Applying a
large pulse on $\varepsilon_\text{e}$ with $\delta \varepsilon_\text{e}
\gg |\Delta_\text{e}|$ which shifts all the levels of the electron dot
levels above the superconducting gap $|\Delta_\text{e}|$ and thereby
empties the electron side of the double dot, we arrive at the state
\begin{equation}\label{eq:ent}
  \Psi_\text{ent} \propto 
  |{\downarrow}_\rs{h} \rangle |+ \rangle + |{\uparrow}_\rs{h} \rangle |-
  \rangle,
\end{equation}
where the spin degree of the last remaining hole is entangled with the
photon polarization; note that alternatively, one could empty the hole
side of the dot to obtain a single electron whose spin is entangled with
the photon polarization.  A drawback of this procedure arises from the
fact that the pulse which empties the dot has to be applied after the
first photon has been emitted and before the second emission takes place.
However, the process is stochastic and measurement is not an option as it
would destroy entanglement.  Therefore, the best we can do is to optimize
the time at which we apply the pulse such as to maximize the probability
that one photon is emitted. We note that the probabilities $P_n(t)$ that
$n$-photons have been emitted at time $t$ to follow the rate equations
\begin{eqnarray}\label{eq:rate_eq}
  \dot P_0 &=& - \Gamma_\text{ph} P_0 \\
  \dot P_1 &=& \Gamma_\text{ph} (P_0 - P_1) \\
  \dot P_2 &=& \Gamma_\text{ph} P_1,
\end{eqnarray}
with the solution $P_0 = \exp(-\Gamma_\text{ph} t)$, $P_1 = \Gamma_\text{ph}
t \exp(-\Gamma_\text{ph} t)$, and $P_2 = 1- P_0 - P_1$. The probability
$P_1$ for a single photon is maximized at a time $t_\text{opt} =
\Gamma_\text{ph}^{-1}$ with the maximal probability $P_1(t_\text{opt})
= 1/e \approx 0.37$ to have a single photon emitted. Conditioning the
experiment on the fact that there is at least a photon emitted, offers
a way to increase the success probability $P_\text{succ}$ to a value
$P_\text{succ}= P_1/(P_1+P_2)= 1/(e-1) \approx 0.58$ for the optimal time
$t_\text{opt}$. In fact, the conditional success probability $P_\text{succ}$
approaches one for short times $t$, i.e, when the kick-out pulse is applied
immediately after the biexciton state has been prepared. However, the large
success probability comes with the expense that the probability $P_1 +
P_2$ to obtain a photon at all becomes vanishingly small.

\section{Spin measurement}

\begin{figure}
  \centering
  \includegraphics{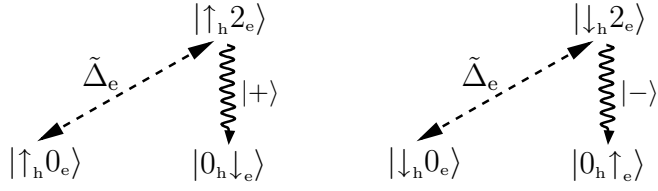}
  \caption{%
  The measurement of the spin of a single hole is done via its conversion
  into the polarization of a photon.  The figure shows the possible initial
  configurations of the hole spin and the production of the photon. The
  latter requires mixing due to a superconducting lead.
  }\label{fig:meas}
\end{figure}

In the situation where we the dots are in state $|1_\rs{h} 0_\rs{e} \rangle$,
we might be interested to find out whether the single hole is in the spin
up or down state.  For example, in the previous section we have discussed
a possible way to generate entanglement between the hole spin and the
polarization of an emitted photon.  In this case, we need to be able to
measure the spin degree of freedom in order to test the entanglement.
We propose a way to transfer the spin state onto the polarization of a
photon which can then be easily probed using a polarizer and a photon
counter; note that this procedure has to be applied fast compared to
$\tilde \Gamma_\text{ph}$ as the secondary photon processes offer a way
to recombine the hole via the creation of a quasiparticle in the lead.
Imagine that the dot is in the state $|{\uparrow}_\rs{h} 0_\rs{e} \rangle$.
By tuning the pair of states $|1_\rs{h} 0_\rs{e} \rangle, |1_\rs{h} 2_\rs{e}
\rangle$ into degeneracy (by choosing $\varepsilon_\text{e} = -U_\text{e} -
U_\text{he}$), we start mixing them into each other with amplitude $\tilde
\Delta_\text{e}$, cf.\ figure~\ref{fig:photon}.  Starting with the state
$|{\uparrow}_\rs{h} 0_\rs{e} \rangle$, we coherently evolve into the state
$|{\uparrow}_\rs{h} 2_\rs{e} \rangle$.  Subsequently, a photon with $+$
polarization can be created and the dot ends up in the state $|0_\rs{h}
{\downarrow}_\rs{e} \rangle$ where it remains until a secondary photon
process occurs.  It is easy to see that if the dot is initially in the
$|{\downarrow}_\rs{h} 0_\rs{e} \rangle$ state the photon produced will carry
the $-$ polarization.  Therefore, we have obtained the situation where the
spin of the hole is transferred into the polarization of a photon thereby
providing a way to measure the spin of the hole.  The same procedure can
also be applied to measure the spin of a single electron if one brings the
states $|0_\rs{h} 1_\rs{e} \rangle$ and $|2_\rs{h} 1_\rs{e} \rangle$ into
degeneracy by choosing $\varepsilon_\text{h} = -U_\text{h} - U_\text{he}$.

\section{Summary}

We have outlined possibilities to use the Josephson LED as device for
quantum information purposes. We have shown the possibility to create
entangled photon pairs on-demand. Furthermore, the device emulates a
two qubit system for which we have proposed a scheme for preparation,
operation, and measurement. We have demonstrated the possibility to
entangle the spin of a particle in one of the dots with the polarization
of an emitted photon. Additionally, we have shown an alternative way to
transfer the spin of the particle into the polarization of a photon which
can be used as a method to measure the spin.

\ack We acknowledge fruitful discussions with Nika Akopian and financial
support from the Dutch Science Foundation NWO/FOM.

\end{document}